# Noise Performance Comparison of 1.5 μm Correlated Photon Pair Generation in Different Fibers

**Qiang Zhou[1, 2], Wei Zhang[1, *], Jie-rong Cheng[1], Yi-dong Huang[1], and Jiang-de Peng[1]**

[1]*Department of Electronic Engineering, Tsinghua University, Beijing, 100084, P. R. China*
[2]*q-zhou06@mails.tsinghua.edu.cn*
[*]*zwei@mail.tsinghua.edu.cn*

**Abstract:** In this paper, the noise performances of 1.5 μm correlated photon pair generations based on spontaneous four wave-mixing in three types of fibers, i.e., dispersion shifted fiber, highly nonlinear fiber, and highly nonlinear microstructure fiber are investigated experimentally. Result of the comparison shows that highly nonlinear microstructure fiber has the lowest Raman noise photon generation rate among the three types of fibers while correlated photon pair generation rate is the same. Theoretical analysis shows that the noise performance is determined by the nonlinear index and Raman response of the material in fiber core. The Raman response raises with increasing doping level, however, the nonlinear index is almost unchanged with it. As a result, highly nonlinear microstructure fiber with pure silica core has the best noise performance and has great potential in practical sources of correlated photon pairs and heralded single photons.

---

## References and links

---

## 1. Introduction

Correlated photon pairs at 1.5 μm have important applications in quantum communication and quantum information processing [1, 2], over large geographic scale, since they are at low loss transmission window of commercial silica fibers. In recent years, correlated photon pair generation through spontaneous four-wave mixing (SFWM) process in optical fibers focuses much attention as a promising way to realizing efficient and compact all-fiber 1.5 μm correlated photon pair sources. However, accompanying with photon pair generation, noise photons would be generated by spontaneous Raman scattering (SpRS), leading to severe deterioration on the performance of 1.5 μm correlated photon-pair generation in fibers. Hence, how to reduce the SpRS noise photons is the key for realizing high quality fiber based 1.5 μm correlated photon pair sources.

According to intensive previous researches on fiber Raman amplification used in optical fiber communications, it is well known that $GeO_2$ concentration in the fibers changes the Raman response of the fiber, which results from the bond-bending motion of the bridging *O* in the $M$-$O$-$M'$ bond ($M$ = Si, Ge) [3]. Hence, in fiber based correlated photon pair generation, the property of generated noise photons by SpRS should be different in fibers with different $GeO_2$ doping levels. Although several types of fibers have been reported to realizing correlated photon pair generation, such as dispersion shifted fiber (DSF) [4-13], traditional single mode fiber [14], microstructure fiber [15-23], and so on, their noise performances have not been compared experimentally. In this paper, the noise performances of correlated photon pair generation in DSF, traditional highly nonlinear fiber (HNLF) and highly nonlinear microstructure fiber (HNMSF) are investigated experimentally. The comparison results shows that microstructure fiber with pure silica core has a better noise performance, which is preferred for practical all-fiber correlated photon pair source at 1.5 μm.

## 2. Experiment on correlated photon pair generation in different fibers

The experimental setup for fiber based correlated photon pair generation is shown in Fig. 1. The pulsed pump light is generated from a passive mode locked fiber laser. A filter system based on a fiber Bragg grating, a circulator and a tunable optical band-pass filter (TOBF) (Dicon, TF-1550-0.8-9) is used to extend its pulse width by narrowing its spectrum. Then it is amplified by an erbium dropped fiber amplifier (EDFA). Amplified spontaneous emission of the EDFA is suppressed by another filter system based on two tunable optical band-pass filters, a fiber Bragg grating and a circulator, achieving a side-band rejection of 115 dB at wavelengths where the signal and idler photon detection is performed. The central wavelength, line width, and repetition rate of the pulsed pump light are 1552.75 nm, 0.2 nm, and 1 MHz, respectively. The duration of the pump pulse is several tens of pico-seconds, estimated by the line width of pump. Before launching the pump pulses into the fiber, a variable optical attenuator (VOA1) and a fiber coupler with a power meter (PM) are used to control and monitor the pump level, while, a polarizer (P) followed by a polarization controller (PC) is used to control its polarization state. Correlated photon pairs are generated by SFWM, accompanied by the noise photons generated by SpRS process, when pump pulses pass through the fiber. The output photons are separated into two parts by a 5/95 fiber coupler. The photons in the 95 percent port are directed into two single photon detectors (SPDs, Id Quantique, id201), through a filtering and splitting system based on a fiber Bragg grating, a 100GHz/40-channels arrayed waveguide grating (AWG, Scion Photonics Inc.), and two tunable optical band-pass filters. Central wavelength and spectral width of the selected signal photons are 1555.15 nm and 0.37 nm, respectively, while 1550.35 nm and 0.37 nm for idler ones, respectively. Total pump isolation is greater than 110 dB at either signal or idler wavelength. The two single photon detectors are operated in gated Geiger mode with a 2.5 ns detection window, and triggered by the pump pulses from the 5 percent port of the coupler detected by a photon detector (PD).

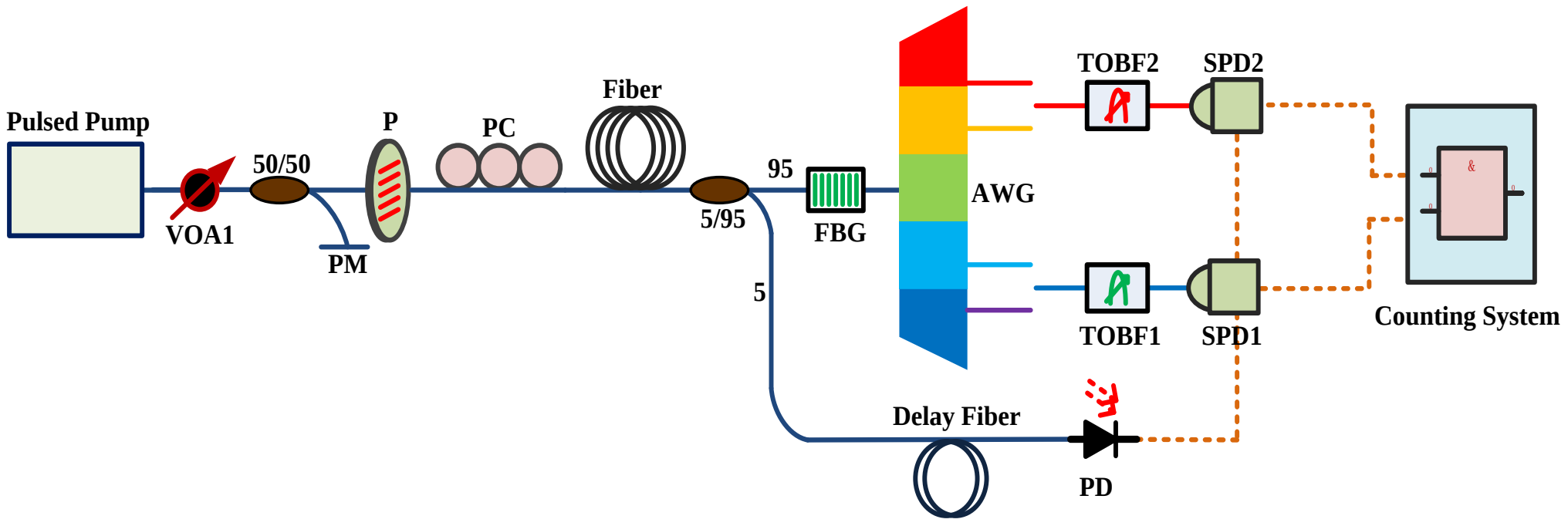

Fig. 1. Experimental setup. VOA, variable optical attenuator; P, polarizer; PM, power meter; PC, polarization controller; FBG, fiber Bragg grating; AWG, arrayed waveguide grating; TOBF, tunable optical band-pass filter; PD, photon detector; SPD, single photon detector.

In the experiment, three types of fibers are used to generate correlated photon pairs, including DSF (G. 653 fiber fabricated by Yangtze Co., Ltd.), traditional HNLF (fabricated by OFS Inc.) and HNMSF (fabricated by Crystal fiber A/S Inc.), respectively. The HNMSF has triangle hole array in cross-section, the diameters of the silica core and air holes are 1.8 μm and 0.89 μm, respectively. Parameters of the three types of fibers related to correlated photon pair generation are listed in table 1.

**Table 1. Parameters of three types of fibers used in the experiment**

| | DSF | Traditional HNLF | HNMSF |
|---|---|---|---|
| Optical nonlinear coefficient | 3 /W/km | 11.1 /W/km | 66.7 /W/km |
| Phase birefringence | -- | -- | $3.5\times10^{-5}$ |
| Dispersion at the pump wavelength | 0.05 ps/km/nm | 0.076 ps/km/nm | 8.65 ps/km/nm |
| Zero dispersion wavelength | 1549 nm | 1548 nm | 1564.1 nm |
| Length | 1000 m | 500 m | 25 m |

Four photon count rates are measured in the experiment. The single-side photon count rates of the signal and idler photons are denoted by $N_s$ and $N_i$, respectively. $N_{co}$ denotes the coincident count rate, representing the result that both single photon detectors record photons simultaneously and the detected photons are generated by the same pump pulse. $N_{ac}$ denotes the accidental coincident count rate, representing the result that both single photon detectors record photons simultaneously, but they are generated by different pump pulses. $N_{co}$ and $N_{ac}$ can be measured by a counting system with a coincident logic circuit. All the experimental data is obtained and averaged under a count time of 30 seconds.

Firstly, correlated photon pair generation is demonstrated under different pump power level using DSF. Fig. 2 (a) and (b) show the measured single-side photon count rates at the signal and idler side. A second-order polynomial, $N_s$, $N_i=s_1N_p+ s_2N_p^2$, is used to fit the experimental data, in which $s_1$ and $s_2$ are the linear and quadratic fitting coefficients, $N_p$ is the pump photon number per pulse, the quadratic term (dashed line) represents the contribution of the photon generated through SFWM process, while the linear term (solid line) represents the contribution of the noise photon generated through SpRS process. Experimental results shows $s_1$=5.315 and $s_2$=6.485 for the signal side, while $s_1$=3.82 and $s_2$=4.13 for the idler side, showing that photons detected in either side have the contribution of SFWM process. Fig. 2 (c) gives the coincident and accidental coincident count rates with increasing signal side photon count rates. It can be seen that the coincident count rates (square dots) are obviously higher than the accidental coincident rates (circular dots), demonstrating the quantum correlation of generated signal and idler photons. The quantum correlated property of the generated photons in the experimental setup is also demonstrated by observing the ratios between $N_{co}$ and $N_{ac}$ under a fixed idler side wavelength (1550.35nm) and different signal side wavelengths, which can be realized easily by selecting different signal side channels of the arrayed waveguide grating. Fig. 2 (d) shows the experimental results. A ratio of 4.24 is achieved under a signal side wavelength of 1555.15 nm, while, the ratios are about 1 under other signal side wavelengths, showing the correlation in wavelength of the generated photon pairs. Similar experimental results also can be obtained using traditional HNLF and HNMSF, showing that correlated photon pairs can be generated and detected successfully in the experimental setup using all the three types of fibers.

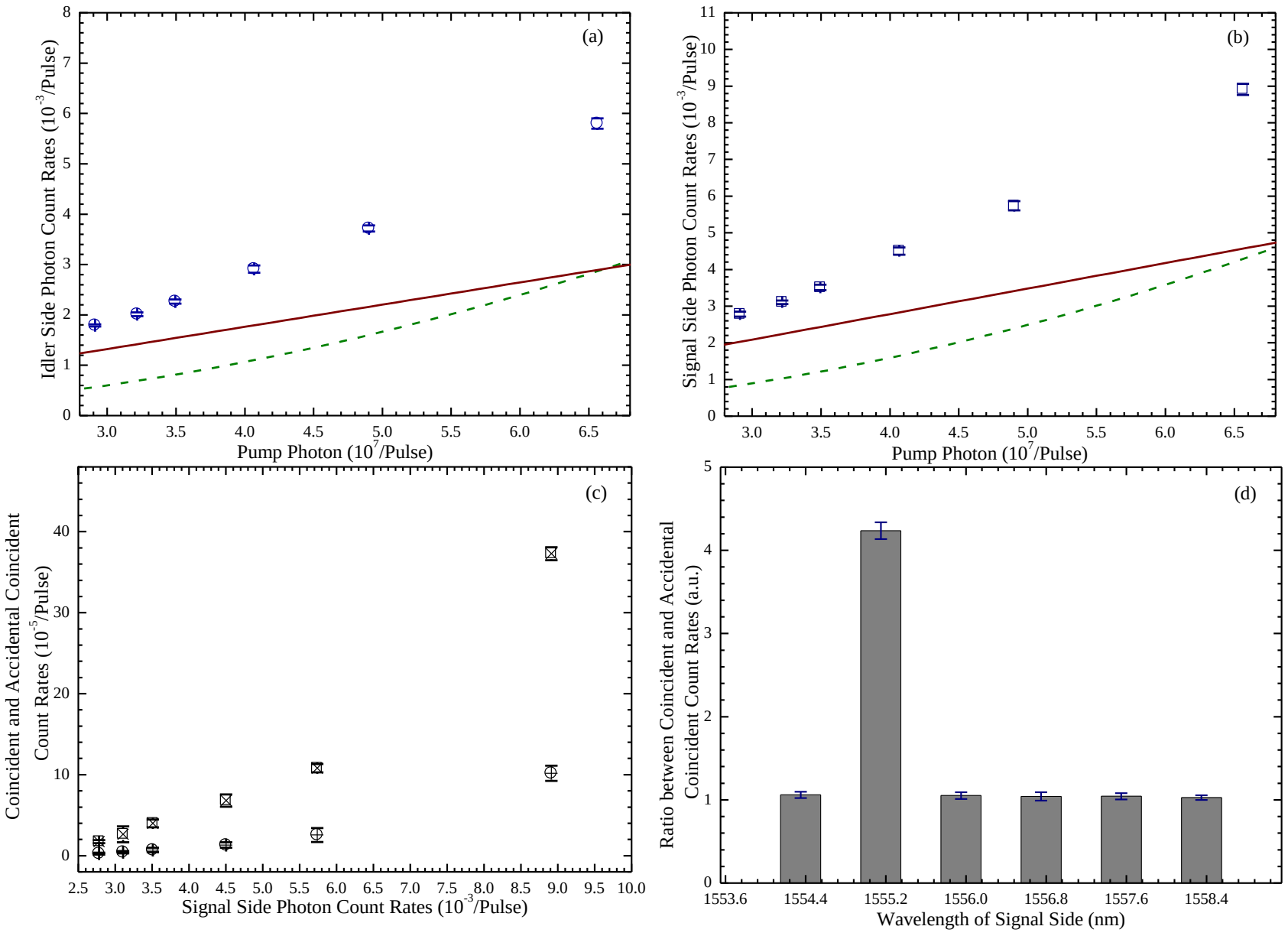

Fig. 2. Experimental results of correlated photon pair generation and quantum correlation property of the generated photons in DSF. Fig. 2 (a), (b) the single side count rates of idler and signal side, respectively; Fig. 2 (c) the coincident (square dots) and accidental coincident (circular dots) count rates under increasing signal side count rate; Fig. 2 (d) the ratio between coincident and accidental coincident counts under six different signal side wavelengths.

The relatively low ratio between $\boldsymbol{N_{co}}$ and $\boldsymbol{N_{ac}}$ is due to the non-correlated noise photons generated by SpRS. In order to investigate the impacts of SpRS noise photons on the correlated photon pair in different fibers, real generation rates of correlated photon pair, stokes and anti-stokes SpRS photons should be obtained. In the experiment, the four photon count rates can be expressed as

$$\begin{aligned}
N_s &= \eta_s(R+R_s)+d_s \\
N_i &= \eta_i(R+R_i)+d_i \\
N_{co} &= \eta_s\eta_i(R+RR_s+RR_i+R_sR_i)+\eta_s(R+R_s)d_i+\eta_i(R+R_i)d_s \\
N_{ac} &= \eta_s\eta_i(R^2+RR_s+RR_i+R_sR_i)+\eta_s(R+R_s)d_i+\eta_i(R+R_i)d_s .
\end{aligned} \tag{1}$$

Where, $\boldsymbol{R}$ is the generation rate of the correlated photon pairs by SFWM process. $\boldsymbol{R_s}$ and $\boldsymbol{R_i}$ is the generation rates of stokes and anti-stokes SpRS photons at signal and idler wavelengths, respectively. $\boldsymbol{d_s}$ and $\boldsymbol{d_i}$ are the dark count rates of SPD1 and SPD2, respectively. The measured photon count rates are also impacted by the collection efficiencies of signal and idler photons, which are determined by losses of the filtering and splitting system and detection efficiencies of the two SPDs. The collection efficiencies are denoted by $\boldsymbol{\eta_s}$ and $\boldsymbol{\eta_i}$ for the signal and idler photons, respectively. Since the dark counting rates and the collection efficiencies can be measured during the experimental setup preparation, $\boldsymbol{R}$, $\boldsymbol{R_s}$, and $\boldsymbol{R_i}$ can be obtained from the experimental data of the four photon count rates according to Eq. (1).

In the experiment, the four photon count rates using DSF, traditional HNLF and HNMSF under increasing pump level are measured, respectively, then, $\boldsymbol{R}$, $\boldsymbol{R_s}$, and $\boldsymbol{R_i}$ in different fibers are calculated according to the experimental data and Eq. (1). It is worth to noting that since the splicing losses between single mode fiber and the three types of fiber used in the experiment are different, $\boldsymbol{\eta_s}$ and $\boldsymbol{\eta_i}$ are collimated before the experiment for each type of fiber. On the other hand, our previous work shows that correlated photon pair generation in HNMSF is influenced by its birefringence [23]. Hence, in the experiment the polarization direction of the pump light is set to one of the polarization axes of the HNMSF by adjusting PC1 to avoid the influence of fiber birefringence.

Fig. 3 is the calculated $\boldsymbol{R}$, $\boldsymbol{R_s}$ and $\boldsymbol{R_i}$ under increasing pump level in DSF, traditional HNLF and HNMSF, shown in figure (a), (b) and (c), respectively. The square, circular, and triangular dots are the experimental results of $\boldsymbol{R}$, $\boldsymbol{R_s}$ and $\boldsymbol{R_i}$, respectively, while, the solid line, dash-dotted line, and dashed line are the fitting curve of them, respectively. As shown in Fig. 3, $\boldsymbol{R}$ increases with pump level quadratically, while, $\boldsymbol{R_s}$ and $\boldsymbol{R_i}$ increase with pump level linearly.

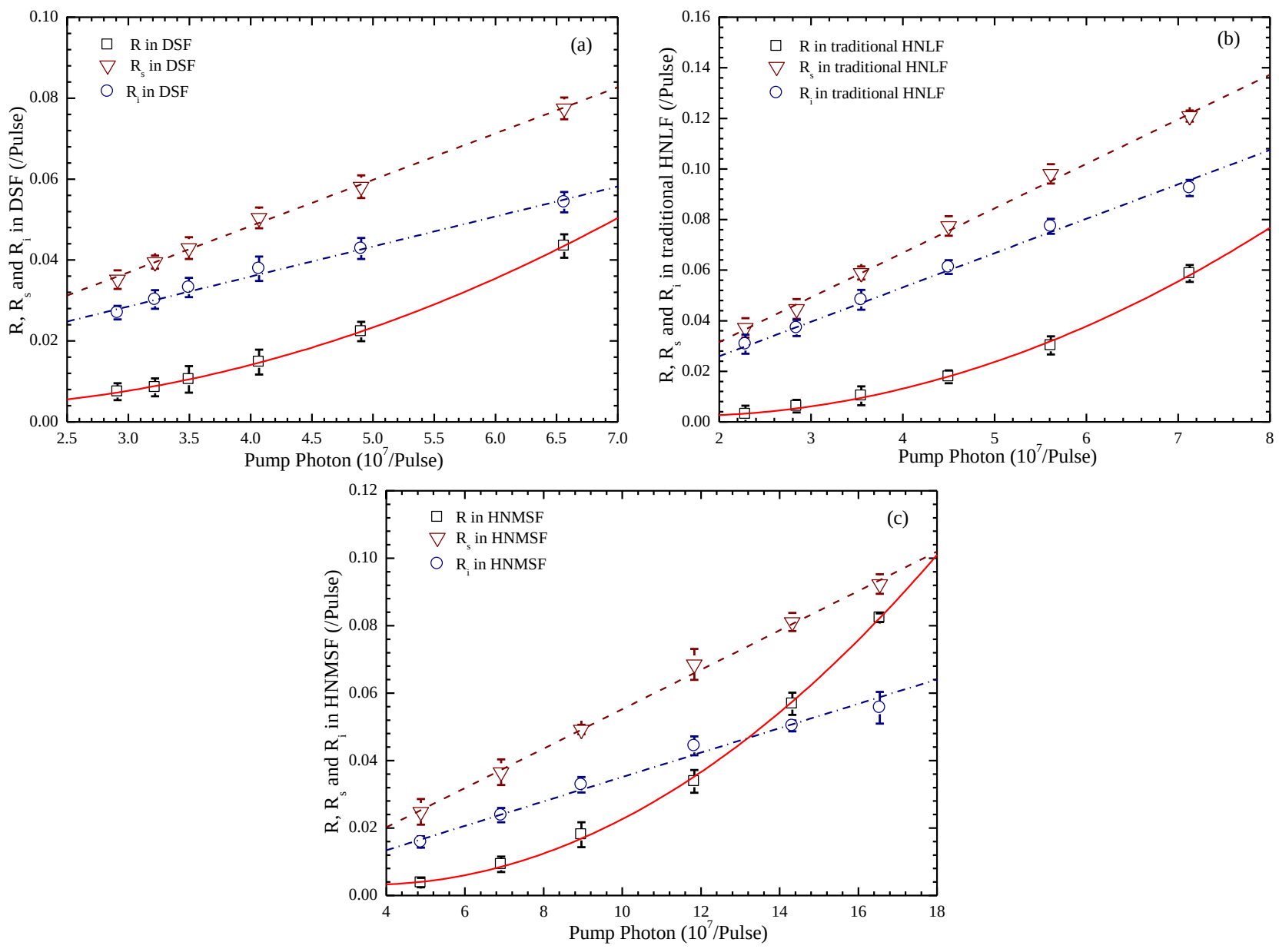

Fig. 3. $\boldsymbol{R}$, $\boldsymbol{R_s}$ and $\boldsymbol{R_i}$ under increasing pump level in DSF (Fig. 3 (a)), traditional HNLF (Fig. 3 (b)) and HNMSF (Fig. 3 (c)). The square, circular, and triangular dots are experimental results of $\boldsymbol{R}$, $\boldsymbol{R_s}$ and $\boldsymbol{R_i}$ respectively. The solid line, dashed-dot line, and dashed line are the fitting curves of them.

It can be seen that SpRS noise photons have large contributions to the generated photons in all the three types of fibers in room temperature under a correlated photon pair generation rate of 0.01~0.1/pulse, which ensures sufficient low multi-photon pair possibility in one pulse. Hence, it can be expected that SpRS noise photons will largely worsen the performance of applications of fiber based correlated photon pair sources. For example, if they are used as heralded single photon sources (HSPSs), in which the idler side photons are detected by single photon detector, providing electrical triggers for the arrival of signal side photons. The preparation efficiency of fiber based HSPSs is determined by the noise performance of correlated photon pair generation and the signal side loss, including component losses insplitting and filtering system and slicing loss between fibers, which can be reduced by proper design of splitting and filtering system, component selection and fiber slicing technique. Hence, the noise performance determines the theoretical up-limit of HSPS preparation efficiency, which is $R/(R+R_i)=1/(1+R_i/R)$, considering that $\boldsymbol{R}$ and $\boldsymbol{R_i}$ are sufficient low and SpRS noise photons do not have quantum correlation characteristics.

To compare the noise performance of correlated photon pair generation in different types of fibers, the ratios between $\boldsymbol{R}$ and $\boldsymbol{R_i}$ in the three types of fibers under increasing $\boldsymbol{R}$ are plotted in Fig. 4. The circular, square and triangular dots are the ratios in HNMSF, DSF, and traditional

HNLF respectively while the solid, dashed and dashed-dot lines are fitting curves using $y = A\sqrt{R}$ , ***A*** is the fitting parameter, which will be discussed in the following section. It can be seen that in all the fibers the ratio between ***R*** and ***R_i*** increases in square root function. The HNMSF shows the best noise performance, while the traditional HNLF shows the worst one. Under the same ***R***, ***R_i*** in HNMSF is about half of the one in traditional HNLF.

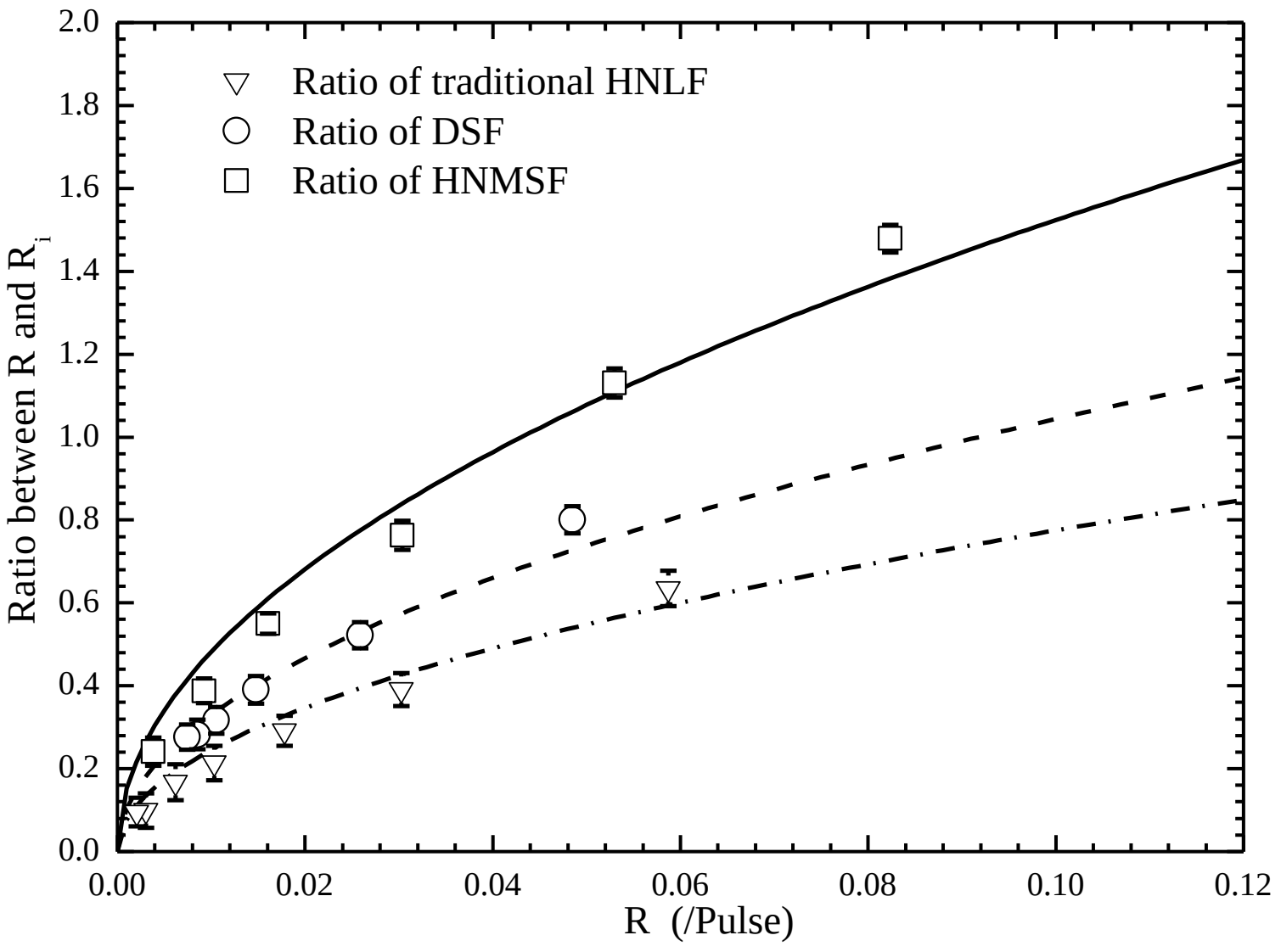

Fig. 4. The ratio between ***R*** and ***R_i***. The circular, square, and triangular dots are the ratio in DSF, HNMSF and traditional HNLF, respectively, while the solid, dashed and dashed-dot lines are the fitting curves of them.

**3. Discussion**

In the experiment, the generation rates of photons (***R***, ***R_s*** and ***R_i***) can be expressed as $\Delta\nu\tau\xi_c(\Omega)\big|_{\Omega=0.3THz}$ , $\Delta\nu\tau\xi_s(\Omega)\big|_{\Omega=0.3THz}$ and $\Delta\nu\tau\xi_i(\Omega)\big|_{\Omega=0.3THz}$ , respectively. Where, $\Delta\nu$ and $\tau$ are the bandwidth of filtering and splitting system and the detection window width of SPDs, which are about 50 GHz and several tens of pico-seconds in both signal and idler sides in the experiment, respectively. $\boldsymbol{\Omega}$ is the frequency detune between the pump light and the generated photon, $\boldsymbol{\Omega}$=0.3 THz in the experiment. $\xi_c(\Omega)$, $\xi_s(\Omega)$ and $\xi_i(\Omega)$ are the photon flux spectral densities of the correlated photon pairs, anti-stokes SpRS photon, and stokes SpRS photon, respectively. According to the quantum theory of describing photons generation in fibers [24, 25], $\xi_c(\Omega)$, $\xi_s(\Omega)$ and $\xi_i(\Omega)$ generated by SFWM and SpRS under a co-polarized single-pump configuration can be obtained by

$$\begin{aligned}
\xi_c(\Omega) &= |\gamma P_0 L|^2 \sin c^2\left[\left(\beta_2\Omega^2 + 2\gamma P_0\right)\frac{L}{2}\right]^2 \\
\xi_i(\Omega) &= P_0 L g_R(\Omega)\left[\exp\left(\hbar|\Omega|/k_B T\right) - 1\right]^{-1} \\
\xi_s(\Omega) &= P_0 L g_R(\Omega)\left[\left(\exp\left(\hbar|\Omega|/k_B T\right) - 1\right)^{-1} + 1\right].
\end{aligned} \tag{2}$$

Where, $\boldsymbol{P_0}$ is the power of pump light. $\boldsymbol{L}$ is the length of fiber. $\boldsymbol{\beta_2}$ is the group velocity dispersion at the pump wavelength. $\boldsymbol{T}$ is the temperature of the fiber. $\gamma$ and $g_R$ are the fiber nonlinear coefficient and co-polarized Raman gain coefficient, respectively.

From Eq. (2) it can be seen that, the maximum $\xi_c(\Omega)$, which is achieved at the phase-matching condition (i.e., value of the *Sinc* function in Eq. (2) equal to 1), is determined by $\gamma P_0 L$ whatever the fiber type. Fig. 5 shows the $\xi_c(\Omega)$ in the three types of fibers used in the experiment under the typical pumping level $\gamma P_0 L = 0.2$ (**R** equals to 0.1/pulse). The solid, dashed and dashed-dot lines correspond to the traditional HNLF, DSF and HNMSF, respectively. The dot line in the figure indicates the frequency detune of narrow band filters for signal and idler photons, which is about 0.3 THz in the experiment. It shows that the value of $\xi_c(\Omega)$ is almost unchanged in the generation bands of all the fibers, although the bandwidth of correlated photon pair generation under co-polarization process are different due to their difference in fiber dispersion. In the experiment, the selected filter detune of signal and idler photons are in the generation bands of all the three fibers, hence, the impact of phase mismatching can be neglected, leading to quadratic increase of $\xi_c$ with increasing pumping level in all the three fibers. Hence, **R** also increase quadratically with increasing pumping level in all the three fibers, which is demonstrated by the experimental result in Fig. 3.

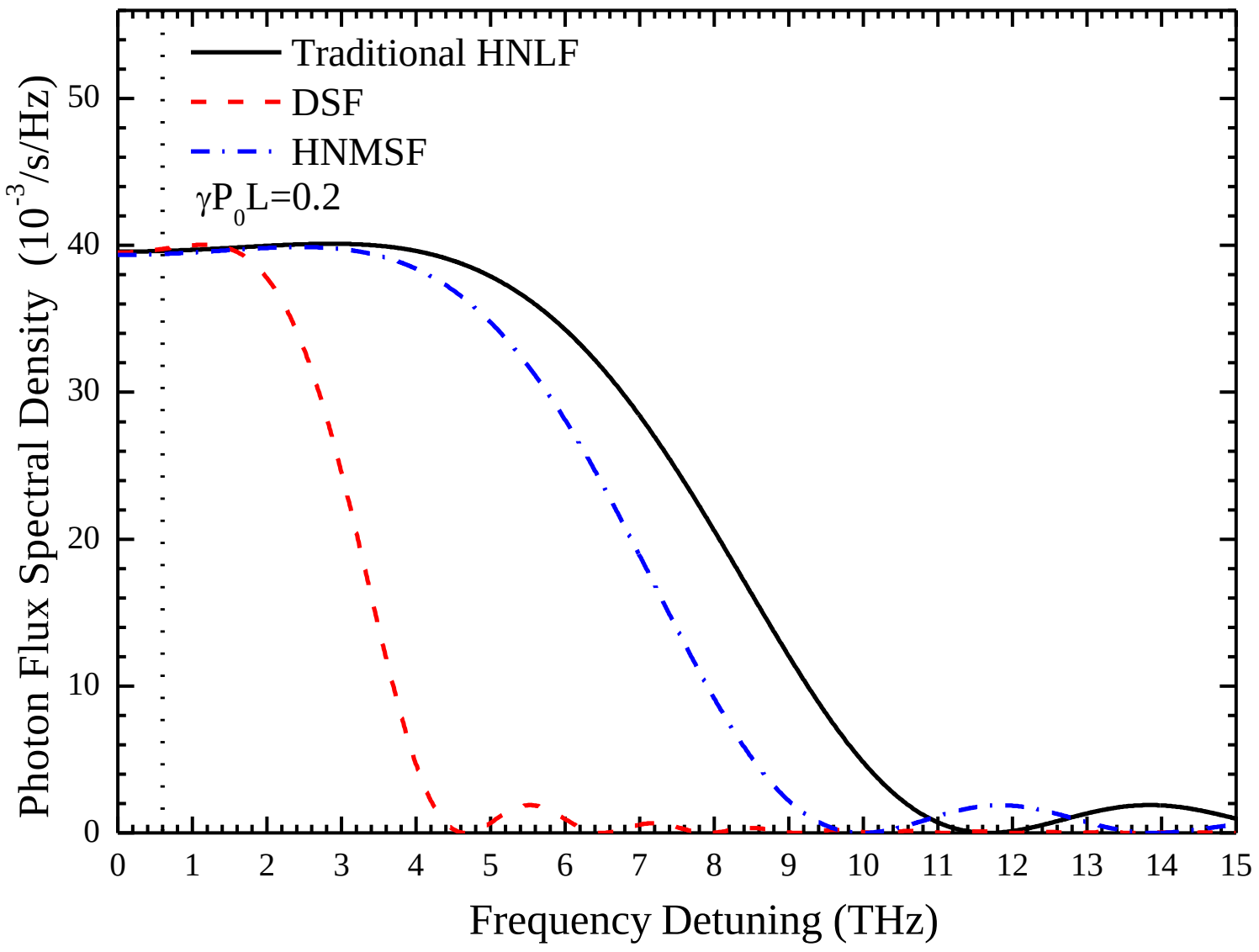

Fig. 5. Photon flux spectral densities of correlated photon pairs in different type fibers.

In order to analysis the noise performance of correlated photon pair generation in different fibers shown in Fig. 4, a simplified expression of the ratio between **R** and $\boldsymbol{R_i}$, which is denoted by $\kappa(\Omega)$, can be deduced by neglecting the impact of phase-matching as

$$\kappa(\Omega)\big|_{\Omega=0.3THz} = \frac{R}{R_i}\Big|_{\Omega=0.3THz} = \frac{\gamma\sqrt{R}}{g_R(\Omega)\left[\exp\left(\hbar|\Omega|/k_B T\right)-1\right]^{-1}\sqrt{\Delta\nu\tau}}\Big|_{\Omega=0.3THz}. \quad (3)$$

Eq. (3) is the origin of the fitting function $y = A\sqrt{R}$ in Fig. 4. The fitting parameter ***A*** should be $\frac{\gamma}{g_R(\Omega)\left[\exp\left(\hbar|\Omega|/k_B T\right)-1\right]^{-1}\sqrt{\Delta\nu\tau}}\Big|_{\Omega=0.3THz}$, which is a constant determined by the fiber properties and fiber temperature. The impact of fiber temperature has been investigated comprehensively in previous works [7, 12]. Our work is focused on the impact of fiber properties.

Eq. (3) indicates that fiber with high nonlinear coefficient and low Raman gain coefficient is preferred for fiber based correlated photon pair generation with low SpRS noise photons. In fibers, $\gamma \propto n_2/A_{eff}$, where $\boldsymbol{n_2}$ is the nonlinear refractive index of the material in fiber core and $\boldsymbol{A_{eff}}$ is the effective area of the fiber. While, $g_R \propto g/A_{eff}$, where ***g*** is the Raman response of it. It can be seen that under the same ***R*** and fiber temperature, $\kappa(\Omega)$ is determined by the nonlinear optical characteristics of the material in fiber core. In traditional fibers, both $\gamma$ and $g_R$ can be improved thanks to the decrease of $\boldsymbol{A_{eff}}$ by high doping. On the other hand, ***g*** increases with the doping level however, $\boldsymbol{n_2}$ is almost unchanged under different doping level [26, 27]. As a result, $g_R$ increases faster than $\gamma$ with increasing doping level. Hence, comparing with DSF, noise performance of the correlated photon pair generation in traditional HNLF is worse due to a higher doping level. On the other hand, HNMSF has pure silica core and its high $\gamma$ is achieved by reducing the core size and extending the air hole fraction. Hence, correlated photon pair generation in HNMSF has the best noise performance among the three types of fibers, which agree well with the experimental results shown in Fig. 4.

To show the impact of fiber property on the performance of applications of fiber based correlated photon pair sources. The theoretical up-limit of the HSPS preparation efficiency is calculated by $R/(R+R_i) = 1/\left(1+1/\kappa(\Omega)\right)\Big|_{\Omega=0.3THz}$, using the fitting curves in Fig. 4 and Eq. (3). Fig. 6 shows the results in the room temperature (300K) and typical temperature achieved by Peltier cooling technique (173K). The gray solid, dashed and dashed-dot lines correspond to the result of HNMSF, DSF and traditional HNLF in 300K, respectively, while, the black solid, dashed and dashed-dot lines correspond to the result of them in 173K, respectively. The dotted line in the figure indicates a typical correlated photon pair generation rate of 0.1/pulse. Under different temperature, ***R*** is almost unchanged and $\boldsymbol{R_i}$ should be modified under different temperature through Eq. (2). It can be seen that under all the conditions (whatever the fiber type and the fiber temperature) the preparation efficiency increases with ***R***, thanks to ***R*** increases faster than $\boldsymbol{R_i}$ with the pump level as shown in the experimental result in Fig. 3. Under a certain fiber temperature and ***R***, HNMSF has the highest preparation efficiency due to its best noise performance among the three types of fibers. It can be seen that, for a typical ***R*** of 0.1/pulse (possibility of multi-photon is 0.45%), the theoretical up-limit of preparation efficiency of the HNMSF based HSPS is 60.37% under room temperature, which is even higher than that based on traditional HNLF under 173K. If the HNMSF is refrigerated to 173K, the theoretical up-limit of preparation efficiency would be 72.85%, which is close to the highest level of HSPS based on correlated photon pair generation by spontaneous parametric down conversion (SPDC) in nonlinear crystal [28]. Hence, HNMSF has great potential in practical sources of correlated photon pairs and heralded single photons, employing proper design of splitting and filtering system, improved fiber splicing technique, and Peltier cooling technique.

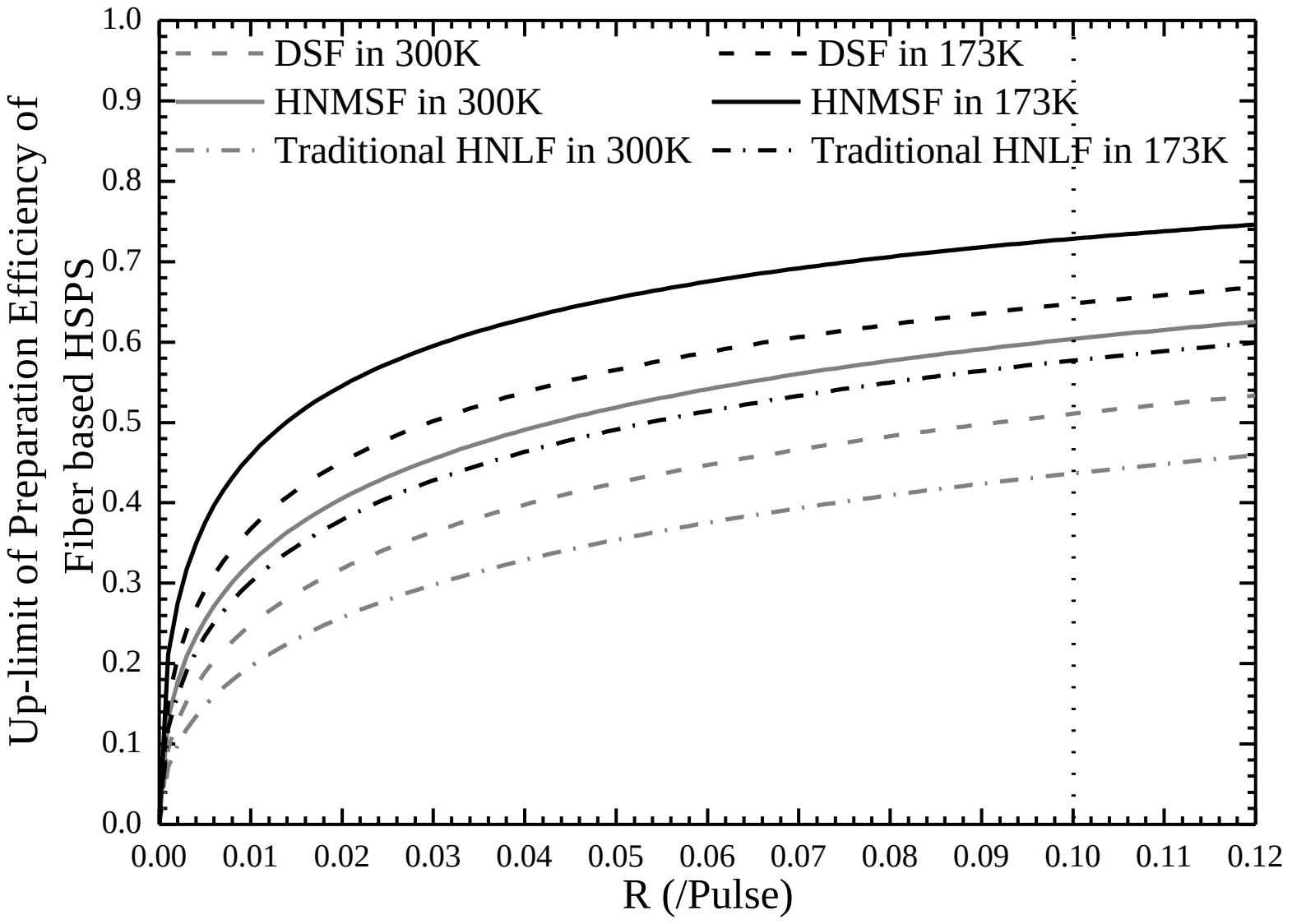

Fig. 6. The up-limit of preparation efficiency of fiber based HSPS under 173K. The solid, dashed and dashed-dot lines correspond to the HNMSF, DSF and traditional HNLF, respectively.

## 4. Conclusion

In this paper, the noise performances of 1.5 μm correlated photon pair generations in DSF, HNLF and HNMSF, are investigated experimentally. Result of comparison shows that HNMSF has the lowest Raman noise photon generation rate among the three types of fibers while correlated photon pair generation rate is the same. Theoretical analysis shows that the noise performance is determined by the nonlinear index and Raman response of the material in the core of fiber. The Raman response raises with increasing doping level, however, the nonlinear index is almost unchanged with it. As a result, HNMSF with pure silica core has the best noise performance and has great potential in practical sources of correlated photon pairs and heralded single photons.

## 5. Acknowledgment

This work is supported in part by National Natural Science Foundation of China under Grant No. 60777032, 973 Programs of China under Contract No. 2010CB327600, Science Foundation of Beijing under Grant No. 4102028, and Basic Research Foundation of Tsinghua National Laboratory for Information Science and Technology (TNList).